\documentclass[epj]{svjour}

\usepackage{hyperref}
\usepackage{graphicx}
\usepackage{amsmath}
\usepackage{amssymb}
\usepackage{microtype}
\usepackage[numbers,sort&compress]{natbib}

\usepackage{soul}
\usepackage[dvipsnames]{xcolor}
\usepackage[switch]{lineno}

\graphicspath{{./Figures}}

\newcommand{\edep}{$E_{\mathrm{dep}}$ }
\newcommand{\edepnospace}{$E_{\mathrm{dep}}$}
\newcommand{\edeprnd}{$E_{\mathrm{dep}}^{\mathrm{rnd}}$ }
\newcommand{\edeprndnospace}{$E_{\mathrm{dep}}^{\mathrm{rnd}}$}
\newcommand{\edepax}{$E_{\mathrm{dep}}^{\mathrm{ax}}$ }
\newcommand{\edepaxnospace}{$E_{\mathrm{dep}}^{\mathrm{ax}}$}

\begin{document}
\emergencystretch 3em

\title{Strong Enhancement of Electromagnetic Shower Development in Oriented Scintillating Crystals and Implications for Particle Detectors}
\titlerunning{Strong Enhancement of Electromagnetic Shower Development in Oriented Scintillating Crystals}

\author{
    Mattia Soldani\inst{1}\thanks{mattia.soldani@lnf.infn.it} \and
    Pietro Monti-Guarnieri\inst{2,3} \and
    Alessia Selmi\inst{4,5} \and
    Nicola Argiolas\inst{6,7} \and
    Luca Bomben\inst{4,5} \and
    Claudia Brizzolari\inst{4,5} \and
    Nicola Canale\inst{8} \and
    Stefano Carsi\inst{4,5} \and
    Nikolaos Charitonidis\inst{12} \and
    Davide De Salvador\inst{6,7} \and
    Vincenzo Guidi\inst{8,9} \and
    Viktar Haurylavets\inst{10} \and
    Mikhail Korzhik\inst{10} \and
    Giulia Lezzani\inst{4,5} \and
    Alexander Lobko\inst{10} \and
    Lorenzo Malagutti\inst{8} \and
    Sofia Mangiacavalli\inst{4,5} \and
    Valerio Mascagna\inst{11} \and
    Andrea Mazzolari\inst{8,9} \and
    Vitaly Mechinsky\inst{10} \and
    Matthew Moulson\inst{1} \and
    Riccardo Negrello\inst{8,9} \and
    Gianfranco Patern\`o\inst{8} \and
    Leonardo Perna\inst{4,5} \and
    Christian Petroselli\inst{4,5} \and
    Michela Prest\inst{4,5} \and
    Marco Romagnoni\inst{8,9} \and
    Federico Ronchetti\inst{13} \and
    Giosu\'e Saibene\inst{4,5} \and
    Francesco Sgarbossa\inst{6,7} \and
    Alexei Sytov\inst{8} \and
    Viktor Tikhomirov\inst{10} \and
    Erik Vallazza\inst{5,3} \and
    Laura Bandiera\inst{8}\thanks{bandiera@fe.infn.it}
}

\institute{
    INFN Laboratori Nazionali di Frascati, Frascati, Italy \and
    Università degli Studi di Trieste, Trieste, Italy \and
    INFN Sezione di Trieste, Trieste, Italy \and
    Università degli Studi dell'Insubria, Como, Italy \and
    INFN Sezione di Milano Bicocca, Milan, Italy \and
    Università degli Studi di Padova, Padova, Italy \and
    INFN Laboratori Nazionali di Legnaro, Legnaro, Italy \and
    INFN Sezione di Ferrara, Ferrara, Italy \and
    Università degli Studi di Ferrara, Ferrara, Italy \and
    Institute for Nuclear Problems, Belarusian State University, Minsk, Belarus \and
    Università degli Studi di Brescia, Brescia, Italy \and
    CERN, Meyrin, Switzerland \and
    École Polytechnique Fédérale de Lausanne, Lausanne, Switzerland
}


\abstract{    
    A particle traversing a crystal aligned with one of its crystallographic axes experiences a strong electromagnetic field that is constant along the direction of motion over macroscopic distances. For $e^\pm$ and $\gamma$-rays with energies above a few $\mathrm{GeV}$, this field is amplified by the Lorentz boost, to the point of exceeding the Schwinger critical field $\mathcal{E}_0 \sim 1.32 \times 10^{16}~\mathrm{V/cm}$. In this regime, nonlinear quantum-electrodynamical effects occur, such as the enhancement of intense electromagnetic radiation emission and pair production, so that the electromagnetic shower development is accelerated and the effective shower length is reduced compared to amorphous materials. We have investigated this phenomenon in lead tungstate (PbWO$_4$), a high-$Z$ scintillator widely used in particle detection. We have observed a substantial increase in scintillation light at small incidence angles with respect to the main lattice axes. Measurements with $120~\mathrm{GeV}$ electrons and $\gamma$-rays between $5$ and $100~\mathrm{GeV}$ demonstrate up to a threefold increase in energy deposition in oriented samples. These findings challenge the current models of shower development in crystal scintillators and could guide the development of next-generation accelerator- and space-borne detectors.
} 

\maketitle
\section*{Introduction}
Inorganic crystal scintillators are essential in high-energy, nuclear, medical and astroparticle physics, since they provide precise measurements of the energy of electrons, positrons and $\gamma$-rays. In particular, numerous high-energy and astroparticle physics experiments rely on high-resolution electromagnetic calorimeters based on scintillating crystals, in which the energy of the incident particle is measured by detecting the scintillation light resulting from the electromagnetic shower it initiates~\cite{gianotti2003}. At the energy frontier (i.e., from the multi-$\mathrm{GeV}$ range upward), showers extend over several tens of radiation lengths ($X_0$), namely the characteristic scale for energy loss of high-energy electrons and positrons via bremsstrahlung and the mean free path for photon pair production (PP). Therefore, detectors in high-energy physics (HEP) require high-$Z$, dense crystals with a small $X_0$ to fully contain the electromagnetic shower within a compact volume.

The formation of electromagnetic showers in inorganic scintillators is conventionally modeled as occurring in amorphous media, i.e., assuming a random spatial distribution of the material atoms. This approach accounts for bremsstrahlung radiation emission and electron-positron pair production by $e^\pm$ and $\gamma$-rays, respectively, interacting with the Coulomb potential of single atoms. However, these models neglect the significant impact of the crystal lattice and its orientation: specifically, when a particle traverses a crystal along one of the strings of atoms in the lattice (i.e., an axis), it experiences an electromagnetic field that is approximately constant along the string direction, resulting from the coherent sum of the single-atom contributions. This phenomenon fundamentally alters the electromagnetic processes~\cite{baier1998electromagnetic,uggerhoj2005}.
 
At sufficiently high energy, the lattice field in the particle rest frame is Lorentz-boosted~\cite{baier1998electromagnetic,uggerhoj2005,Baryshevskii89,sorensen1996_notes} and can reach an amplitude larger than the Schwinger critical field ($\mathcal{E}_0 = m_e^2 c^3/e \hbar \sim 1.32 \times 10^{16}~\mathrm{V/cm}$), i.e., the threshold for nonlinear quantum-electrodynamical (QED) effects to occur~\cite{schwinger51}. This is the so-called Strong Field (SF) regime~\cite{uggerhoj2005}. Such an intense field induces an enhancement of the quantum radiation emission probability with respect to the Bethe-Heitler description typical of amorphous media~\cite{kimball1985}. By crossing symmetry, the probability for the creation of an $e^\pm$ pair by a high-energy photon is significantly increased as well~\cite{baier1998electromagnetic,uggerhoj2005,Baryshevskii:1983JETP,BARYSHEVSKII1985335}. Radiation emission by $e^\pm$ is enhanced approximately twice as strongly as pair production by photons at the same energy scale \cite{baryshevsky2017}.

The SF regime is attained if $\chi = \gamma \mathcal{E}_\mathrm{lab} / \mathcal{E}_0 \gtrsim 1$, where $\mathcal{E}_\mathrm{lab}$ is the continuous axial electric field in the laboratory frame and $\gamma$ is the Lorentz factor, which translates into an energy threshold for the primary particle~\cite{Baryshevskii89,baier1998electromagnetic,uggerhoj2005}. This threshold is $\mathcal{O}(10~\mathrm{GeV})$ for high-$Z$ materials and is roughly the same for both bremsstrahlung and pair production: for the latter, it can be estimated by replacing $\gamma$ with $\gamma_\mathrm{PP} = \hbar \omega / m_e c^2$, $\hbar \omega$ being the incident photon energy. The intensity of the SF effect grows with $\chi$ up to saturation, which is typically reached far above several $\mathrm{TeV}$, beyond the achievable experimental conditions in current and planned high-energy accelerators~\cite{uggerhoj2005}.

Overall, the SF-related increase of these processes is expected to lead to an enhancement of the electromagnetic shower development, i.e., to a larger production of secondary particles per unit of length, with respect to the case of an amorphous or randomly oriented medium. In the case of a scintillating crystal, this results in a larger light yield (i.e., the number of scintillating photons emitted inside the medium) per unit of shower depth. 

\begin{figure}[htbp!]
    \centering
    \includegraphics[width=\linewidth]{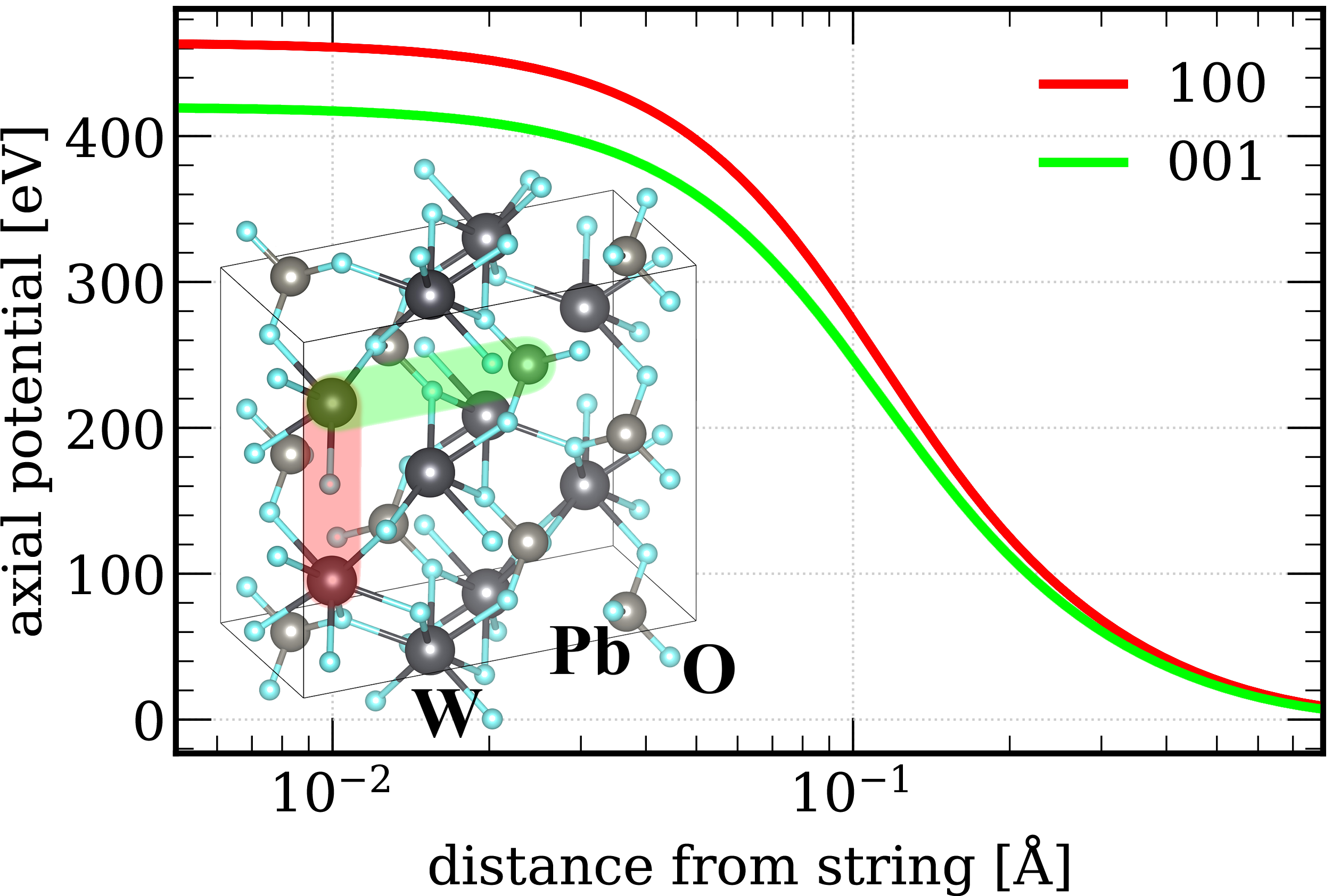}\\
    \caption{Average continuous axial potential felt by $e^{+}$ (opposite sign for $e^{-}$) for two of the main PbWO$_4$ axes as a function of the distance in transverse direction from the atomic string. The corresponding crystal structure is shown in the insert.}
    \label{fig:potentials}
\end{figure}

For instance, let us consider lead tungstate (PbWO$_4$ or PWO, $X_0 \sim 0.89~\mathrm{cm}$~\cite{pdg}), the highest-density inorganic scintillator crystal commonly used in electromagnetic calorimeters, such as those in the CMS experiment at CERN~\cite{CMS_ECAL} and the PANDA experiment at GSI~\cite{PANDA_ECAL}, as well as in the design for the IDEA detector at the Future Circular Collider~\cite{FCC_IDEA}. The average electric potentials associated to two of the strongest (i.e., higher-charge-density) axes of PbWO$_4$, $\langle 100 \rangle$ and $\langle 001 \rangle$, are shown in figure~\ref{fig:potentials}. The behaviour of these potentials as a function of the distance from the atomic string in the transverse plane was calculated with the approach introduced in \cite{baier1998electromagnetic}. In case of both these axes, the full SF regime ($\chi = 1$) is attained at an energy of about $25~\mathrm{GeV}$. The corresponding values of $U_0$, i.e., the average potential well depth~\cite{uggerhoj2005}, are approximately $460~\mathrm{eV}$ and $420~\mathrm{eV}$, respectively~\cite{soldanithesis}.

To assess the influence of the lattice orientation on the development of electromagnetic showers, it is crucial to evaluate how precisely the particle trajectory must be aligned with the crystal axis, for the SF regime to be achieved. In terms of the misalignment angle $\theta_{\rm mis}$, defined as the angle between the direction of incidence and the crystallographic axis, the characteristic scale over which electromagnetic processes are enhanced by strong-field effects is $\Theta_0 = U_0 / m_e c^2$, as described in~\cite{uggerhoj2005}. For PbWO$_4$, the corresponding angular acceptance is approximately $0.9~\mathrm{mrad}$ for both axes. For $\theta_\mathrm{mis} > \Theta_0$, the SF effects are less pronounced, while other coherent effects occur, i.e., coherent bremsstrahlung and coherent pair production~\cite{dyson_anisotropy_1955,diambrini-palazzi_interazioni_1962}. It has been observed that such phenomena give rise to an increase of both bremsstrahlung and pair production for $\theta_\mathrm{mis}$ up to $1^\circ$ ($17~\mathrm{mrad}$) and also for energies significantly smaller than the SF scale~\cite{andersen1983_notes}.

Experimental investigations of the SF effects in both radiation emission~\cite{cue1984} and pair production~\cite{belkacem1984} started in the 1980s. They were initially focused on very thin ($\lesssim 1~\mathrm{mm}$) silicon, germanium and diamond crystals. Studies on high-$Z$, high-density metallic materials, such as iridium and tungsten, were performed starting from the 1990s, with the aim of developing compact photon converters and high-intensity positron sources~\cite{moore1996,kirsebom1998,artru2005,bandiera2022,chehab2002, soldani2023, Chaikovska_2022}; in facts, only optically opaque and relatively thin crystals were examined. More recently, the enhancement of radiation from $120~\mathrm{GeV}$ electrons incident on the $\langle 001\rangle$ axis of a $0.45~X_0$ thick PbWO$_4$ crystal was studied~\cite{bandiera2018}: this was the first investigation of the SF-induced ($\chi \gg 1$) enhancement of radiation emission occurring in an oriented inorganic scintillator. No direct measurement of the energy deposited inside the crystal sample was made in that case. The only other study to date on PbWO$_4$ crystals was performed with $26$-$\mathrm{GeV}$ electrons, i.e., approximately at the SF threshold~\cite{baskov1999}: as emphasized by the authors themselves, the results, although promising, were of limited use due to the absence of a physical model with which to compare the data, as well as the need for further investigation with different particles and at higher energies, specifically, in the $\gtrsim 100~\mathrm{GeV}$ range, where the SF regime is fully attained.

In this manuscript, we present a direct measurement of the enhancement of the electromagnetic shower development that occurs in oriented crystals in the full SF regime, for the first time exploring the range of particle energies and crystal thicknesses which is of interest for particle detector development. This achievement was made possible by the use of PbWO$_4$ samples of known lattice orientation. Specifically, we performed experimental tests at the extracted beamlines of the CERN Super Proton Synchrotron (SPS)~\cite{banerjee_north_2021} using beams of high-energy electrons and $\gamma$-rays, measuring the crystal light yield with a photodetection system based on Silicon PhotoMultipliers (SiPMs)~\cite{soldani2022_pwo,selmi2023}. We compared our results to a Monte Carlo model developed specifically for this work, demonstrating an excellent agreement between data and simulations. The results of this study demonstrate that the shower formation, and hence the light yield of oriented scintillating crystals, is significantly enhanced with respect to the non oriented case. This will pave the way towards the development of innovative electromagnetic calorimeters and preshower detectors, and also towards a more accurate understanding of the performance of any existing or planned crystalline detector.

\section*{Materials and methods}
\subsection*{Crystal samples}
Four oriented PbWO$_4$ samples were probed, with thicknesses of $0.45$, $1$, $2$ and $4.6~X_0$, thus covering the initial part of the electromagnetic shower development, in which the most pronounced enhancement is expected. These four thickness values were chosen to cover three ranges of interest, specifically, the $< 1~X_0$ range, to isolate the first interactions in the shower, and the $1$--$2~X_0$ and $4$--$5X_0$ ranges, which match the crystals currently considered for applications in many space-borne \cite{Atwood09} and accelerator-based detectors~\cite{IDEA2020, crilin}, respectively. The $4.6~X_0$-thick PbWO$_4$ sample was obtained by cutting a longer crystal from a prototype developed for the endcap part of the CMS ECAL (Electromagnetic CALorimeter)~\cite{CMS_ECAL}, while the others were produced by CRYTUR and MolTech.

Two of the main axes of PbWO$_4$ were studied: the $0.45$ and $4.6~X_0$ crystals were oriented along the $\langle 100 \rangle$ axis, while the $1$ and $2~X_0$ crystals were oriented along the $\langle 001 \rangle$ axis. As anticipated, these two axes feature approximately the same SF characteristics.

Each crystal sample was probed at different values of $\theta_{\text{mis}}$, from the condition of incidence on axis ($\theta_{\text{mis}} = 0~\mathrm{mrad}$) up to that corresponding to random orientation ($\theta_{\text{mis}} \sim 50~\mathrm{mrad} \sim 3^\circ$).

\subsection*{Experimental setup}
Our studies were performed on the H2 extracted beamline at the CERN Super Proton Synchrotron~\cite{banerjee_north_2021} with a $120$-$\mathrm{GeV}/c$ electron beam with $\lesssim 100$~{\textmu}$\mathrm{rad}$ divergence, which allowed us to reach $\chi \sim 5$, where strong SF effects become significantly pronounced. The $1~X_0$ crystal was also probed with a tagged photon beam with an energy of $5$--$100~\mathrm{GeV}$~\cite{pmg2022}. The crystals were installed on a high-precision goniometer, comprising a linear stage, to move the samples along the transverse horizontal axis, and two angular stages, to rotate the samples around the vertical and horizontal transverse axes in order to align them to the beam direction with a precision of $1~${\textmu}$\mathrm{rad}$~\cite{BANDIERA2013135}.

The incident particle trajectories were reconstructed with a tracking system based on silicon microstrip detectors, which provide excellent angular resolution with a minimal amount of material along the beam path. Each tracking module consists of a pair of single-sided silicon tiles of about $9.5 \times 9.5~$$\mathrm{cm}^2$ active area and $410~${\textmu}$\mathrm{m}$ thickness, arranged in an $xy$ scheme. The strips have a readout pitch of $242~${\textmu}$\mathrm{m}$ and feature analog readout, which results in a single-hit spatial resolution of $30~${\textmu}$\mathrm{m}$~\cite{lietti2013}. Using two tracking modules at a distance of about $15~\mathrm{m}$ resulted in an angular resolution on $\theta_{\text{mis}}$ of $\mathcal{O} (1~${\textmu}$\mathrm{rad})$, i.e., about $10^{-3}~\Theta_0$.

The energy of the particles emerging from the crystals was measured by a homogeneous electromagnetic calorimeter consisting of lead glass blocks in different geometric configurations. Each block has a thickness of about $27X_0$ ($37~\mathrm{cm}$) and a trapezoidal shape, with a front (rear) section of approximately $10 \times 10~\mathrm{cm}^2$ ($11 \times 11~\mathrm{cm}^2$)~\cite{opal}.

\begin{figure*}[htbp!]
    \centering
    \includegraphics[width=\linewidth]{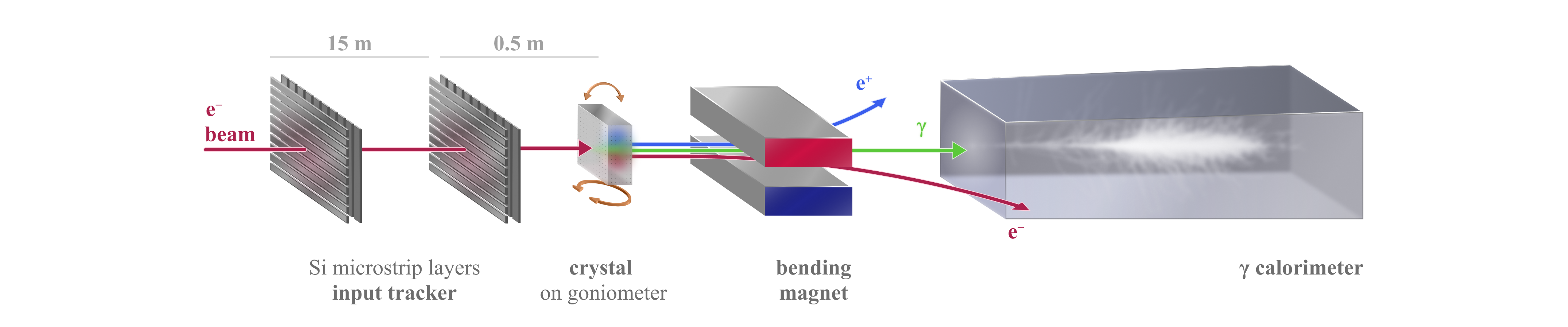}
    \includegraphics[width=\linewidth]{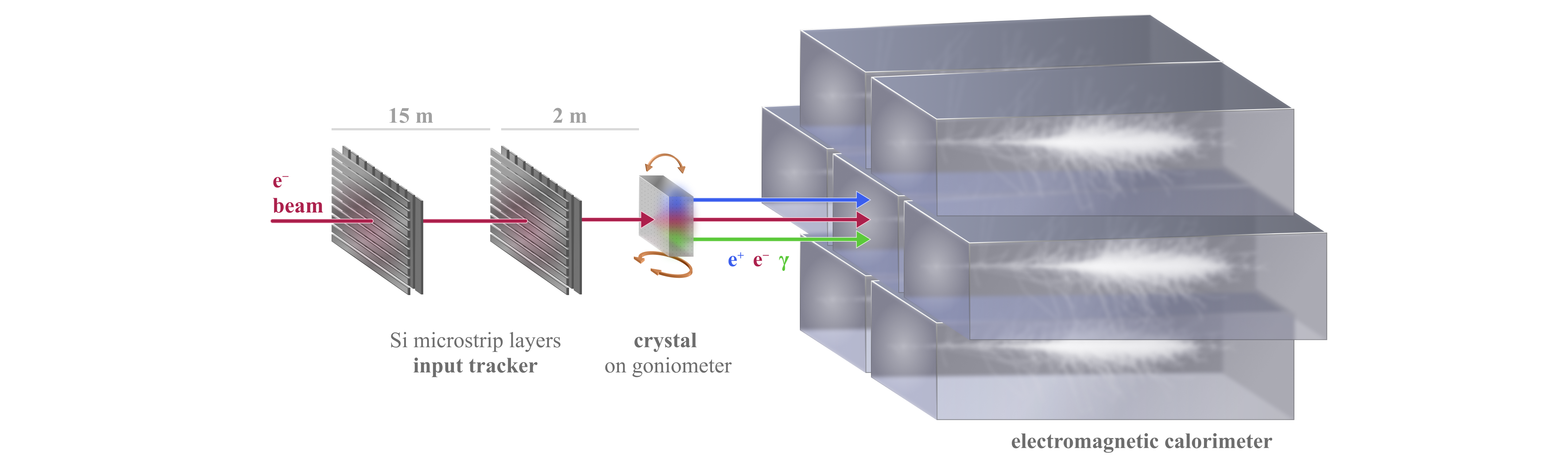}
    \includegraphics[width=\linewidth]{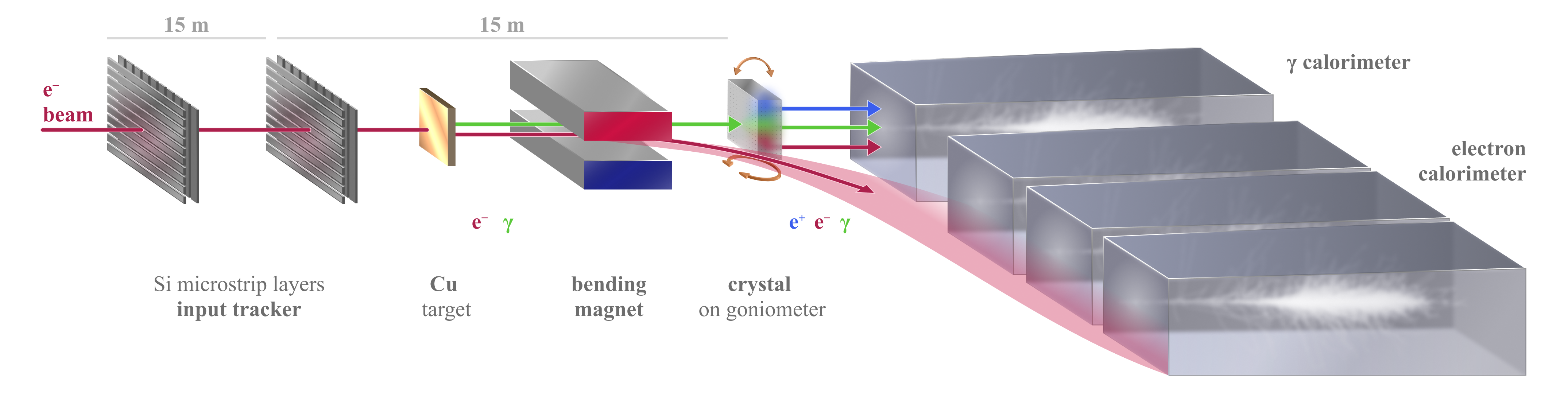}
    \caption{Scheme of the experimental setup in various configurations---optimized for thin crystal alignment \textit{(top)}, for enhanced hermeticity \textit{(center)} and for measurements with tagged photons \textit{(bottom)}.}
    \label{fig:setups}
\end{figure*}

The layout of the experimental setup used in the various configurations is shown in figure \ref{fig:setups}. For the $0.45~X_0$ sample, the layout sketched in figure \ref{fig:setups} (top) was chosen, in which charged particles from showers initiated in the crystal are swept by a bending magnet out of the acceptance of the downstream calorimeter, which consists of a single lead-glass block~\cite{soldanithesis,pmg2023}. As shown in~\cite{bandiera2018}, the difference between the shapes of the energy spectra for forward photons measured in random alignment and on axis provides good contrast between the two angular configurations. For crystals with thickness $< 1~X_0$, this constitutes a much better criterion than the total energy of the particles emerging from the crystal for the determination of the angular position corresponding to axial alignment.

On the other hand, owing to the significant variation of the energy deposited inside the samples, $E_{\mathrm{dep}}$, between different angular configurations, a measurement of the total final-state energy proves useful to validate the angular alignment. Indeed, as shown in figure \ref{fig:lightspectra}, the latter quantities are strongly correlated to each other. This was achieved with the layout sketched in figure \ref{fig:setups} (center): an array of lead glass blocks was positioned at $\sim 30~\mathrm{cm}$ from the rear face of the crystal, which guarantees good hermeticity for detection of all particles emerging from the crystal at angles of up to $\sim 30^\circ$~\cite{soldanithesis}. In this configuration, the total energy measured by the lead glass blocks is $E_{\mathrm{CAL}} \sim 120~\mathrm{GeV} - E_{\mathrm{dep}} - \mathcal{O}(10~\mathrm{GeV})$, where the rightmost term accounts for residual transverse losses and was evaluated with simulations.

The setup used for measurements with tagged photons is shown in figure \ref{fig:setups} (bottom). A copper target with a thickness of $1~\mathrm{mm}$ ($0.07~X_0$) was used to generate bremsstrahlung photons from the incident $120~\mathrm{GeV}$ electrons, which were then diverted by a bending magnet towards an array of lead glass blocks: this spectrometer was used to measure the electron momentum, $p_{\mathrm{e}}$, and, in turn, to estimate the energy of the bremsstrahlung photon as $120~\mathrm{GeV} - p_{\mathrm{e}} c$~\cite{soldani2023,pmg2023}. The latter showed excellent agreement with the energy inferred as the sum of the energy deposited in the crystal and in the lead glass block placed at the crystal rear face ($\gamma$ calorimeter), i.e., $E_\gamma^\mathrm{inf} = E_{\mathrm{dep}} + E_{\mathrm{CAL,\gamma}}$. Overall, the apparatus was sensitive to $E_\gamma^\mathrm{inf}$ between $5$ and $100~\mathrm{GeV}$.

\subsection*{Photodetection system}
The energy deposited inside the samples, \edepnospace, was evaluated from the scintillation light measured with SiPMs. The $0.45~X_0$ crystal was read out by a single {ASD-NUV4S-P} model by AdvanSiD~\cite{soldani2022_pwo}. It has a $4 \times 4~\mathrm{mm}^2$ square surface, which well matches the smaller faces of the sample, and a photodetection efficiency that matches the PbWO$_4$ emission spectrum. The SiPM was coupled to an {ASD-EP-EB-N} evaluation board~\cite{soldani2022_pwo}. All the other samples were coupled to arrays of Onsemi ARRAYC-60035-4P-BGA SiPMs~\cite{selmi2023}.

The $0.45$, $1$ and $2~X_0$ crystals were read out from one of the faces parallel to the beam direction, whereas the $4.6~X_0$ was read out from the rear face. The calibration of the SiPM response into GeV was performed by equalizing the measured random-orientation scintillation peak with the corresponding energy deposit peak from simulations.

\subsection*{Simulation methods}
The on-axis and random data were compared to the results of MC simulations performed using the Geant4 toolkit~\cite{g4}, properly modified to include SF effects. In particular, the custom physics model that we developed is based on the Baier-Katkov quasi-classical operator method~\cite{baier1998electromagnetic}. This method is applied by correcting the differential bremsstrahlung and pair production cross-sections with energy-dependent factors~\cite{bandiera2018,baryshevsky2017} that are computed through full MC simulations of the particle motion inside the crystalline lattice at different energies and angles: here, the radiation emission and pair production probabilities in the axial field are computed by directly integrating the quasi-classical Baier-Katkov formula over realistic particle trajectories~\cite{Guidi12,RADCHARM++,CRYSTALRAD,sytov2023,Bandiera2021}. 

\section*{Results}
\subsection*{Enhancement of the scintillation light yield and energy deposit}
\begin{figure}[t!]
    \centering
    \includegraphics[width=\linewidth]{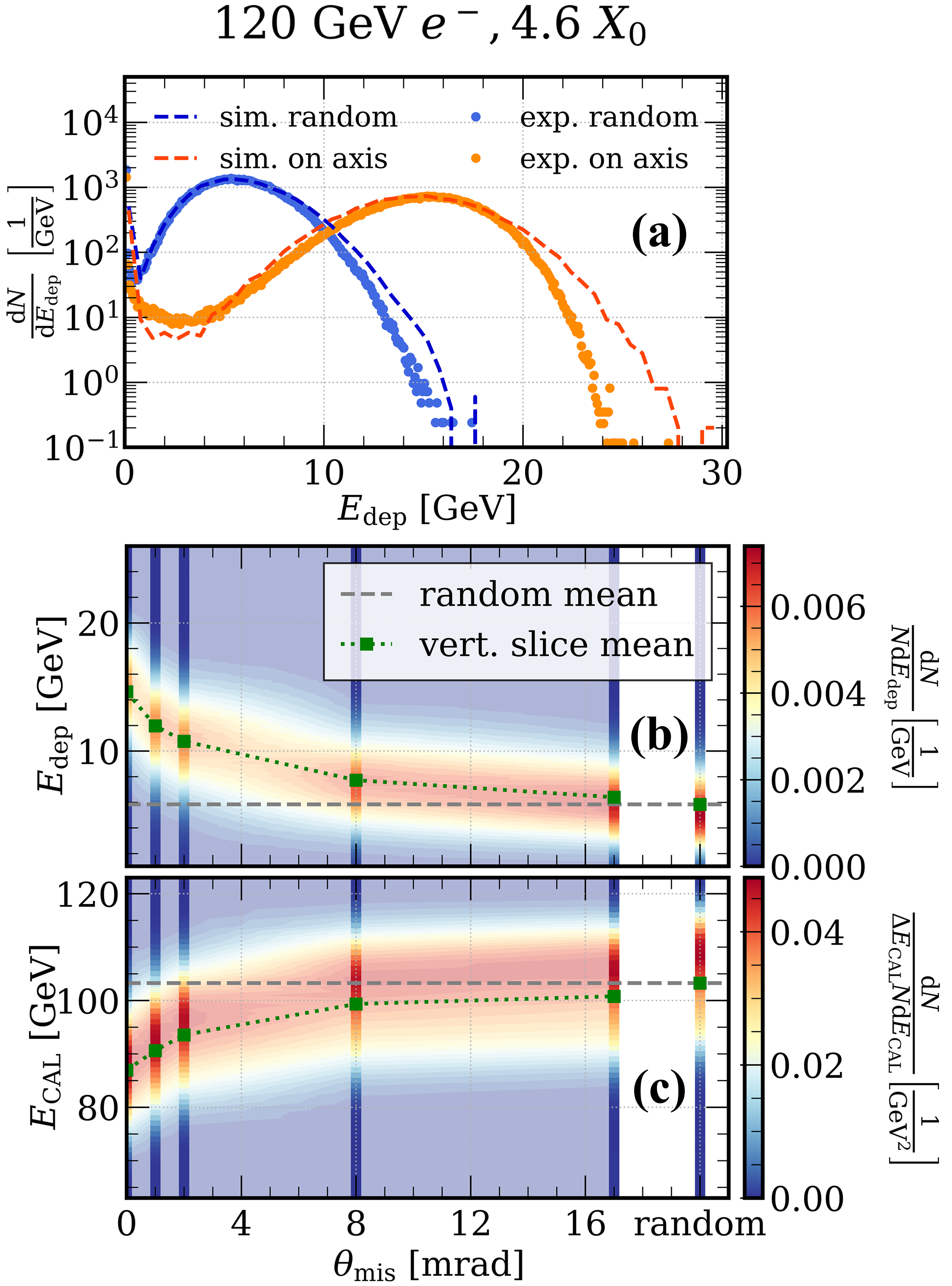}
    \caption{Experimental measurement with $120~\mathrm{GeV}$ electrons in the $4.6~X_0$ PbWO$_4$ sample: \textit{(a)} distribution of the deposited energy, \edepnospace, in the crystal in randomly oriented and axial configurations, and corresponding simulated (dashed) curves; \textit{(b)} \edep as a function of the angle between the beam and the axial direction, $\theta_{\text{mis}}$; \textit{(c)} corresponding energy deposited in the calorimeter positioned downstream of the crystal, $E_{\mathrm{CAL}}$, as a function of $\theta_{\text{mis}}$.
    The vivid part of the contour plot corresponds to the experimental data; the shaded part to a linear interpolation between Delaunay triangles calculated from the available data. The green squares indicate the mean values at different angles. The point on the $x$-axis corresponding to the randomly oriented configuration ($\sim 50~\mathrm{mrad}$) is not to scale.}
    \label{fig:lightspectra}
\end{figure}

In order to estimate the strength of the electromagnetic shower modification, the energy deposited inside the crystal samples, \edepnospace, was evaluated from the scintillation light measured with the SiPMs. Specifically, the calibration of the experimental data into energy units (GeV) was performed by equalizing the scintillation peaks measured in random orientation with the corresponding simulated energy deposit peak. This procedure was repeated separately for each crystal. The distribution of the energy deposited by the $120~\mathrm{GeV}$ electron beam in the $4.6~X_0$ sample (the thickest tested), both in random and on-axis alignment, is shown in figure~\ref{fig:lightspectra}a. Here, the data collected in axial configuration include particle trajectories impinging on the axis with an angle of less than $100$~{\textmu}$\mathrm{rad}$, i.e., the size of the beam divergence, and about $0.1~\Theta_0$. An excellent agreement between the measured and simulated \edep spectra has been found in both alignment conditions.

Figure~\ref{fig:lightspectra}b shows measurements obtained at different values of $\theta_{\text{mis}}$. The energy deposited inside the crystal grows as $\theta_{{\rm mis}}$ decreases: the measured mean \edep is about $5~\mathrm{GeV}$ in random orientation and $15~\mathrm{GeV}$ on axis ($\theta_{\text{mis}} \sim 0$). The enhancement in the energy deposit is particularly pronounced for $\theta_{\text{mis}} \lesssim \Theta_0 \sim 0.9~\mathrm{mrad}$. Furthermore, a smaller enhancement is still observed for $\theta_{\text{mis}}$ as large as $17~\mathrm{mrad}$ ($1^\circ$). Moreover, the energy deposited in the lead-glass calorimeter placed downstream of the crystal samples (denoted as $E_{\mathrm{CAL}}$) increases with $\theta_{\text{mis}}$, as shown in Fig.~\ref{fig:lightspectra}c. 

It is important to note that the combined detection system (crystal and lead-glass calorimetr) is not fully hermetic in the transverse direction. Consequently, a systematic missing energy on the order of $\mathcal{O}(10~\mathrm{GeV})$ is observed, corresponding to the difference between the nominal beam energy and the sum of the energy deposits in the PbWO$_4$ crystal and lead-glass calorimeter, i.e., $120~\mathrm{GeV} - E_{\mathrm{dep}} - E_{\mathrm{CAL}}$. Indeed, this missing energy is attributed to transverse losses from particles exiting the crystal at large angles, outside the geometric acceptance of the calorimeter. This was confirmed by the Geant4 simulations of the exeprimental setup. 


\subsection*{Electrons vs photons}
A second series of tests was carried out on the $1~X_0$ sample, which is of interest as it corresponds to the typical longitudinal segmentation adopted in satellite-borne detectors for very-high-energy (VHE) $\gamma$-ray and cosmic-ray observation~\cite{Atwood09, calet}. Therefore, these tests were conducted using both electrons at $120~\mathrm{GeV}$ and photons in the energy range $5$--$100~\mathrm{GeV}$. The photon beam was bremsstrahlung-tagged, i.e., the energy of each photon, produced via bremsstrahlung from electrons interacting with a radiator, could be determined event-by-event by measuring the energy of the corresponding deflected electron.

\begin{figure*}[htbp!]
    \centering
    \includegraphics[width=\linewidth]{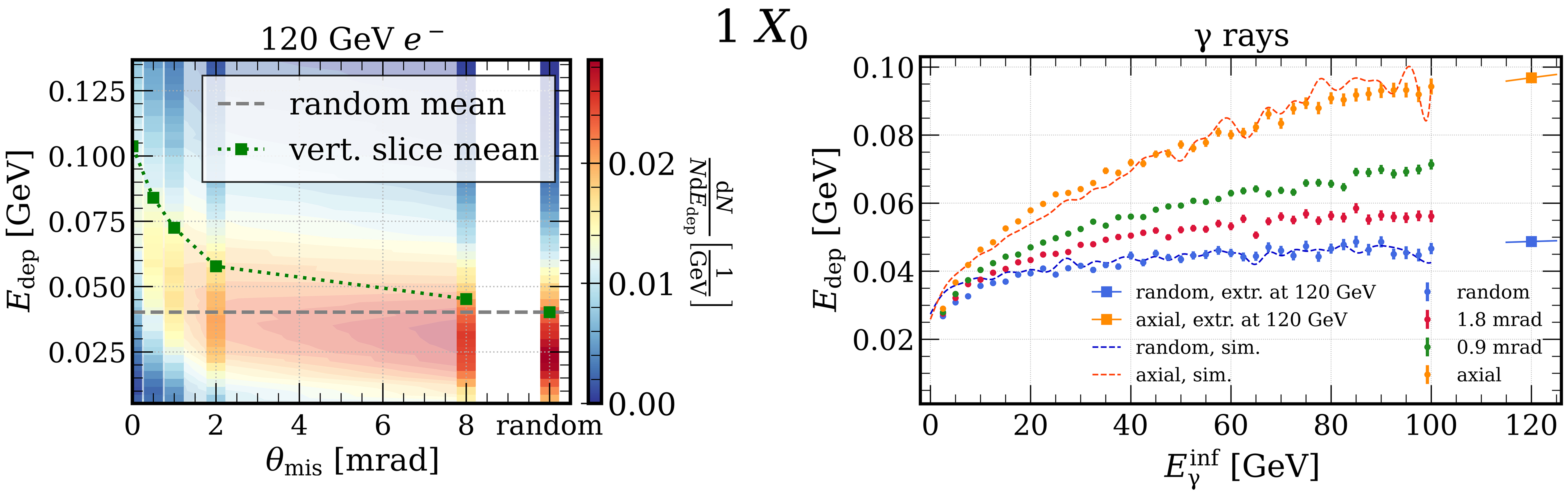}
    \caption{Measurements on the $1~X_0$ sample. \textit{Left}: \edep by $120~\mathrm{GeV}$ electrons as a function of $\theta_{\text{mis}}$. The vivid part of the contour plot corresponds to the experimental data; the shaded part to a linear interpolation between Delaunay triangles calculated from the available data. The green squares indicate the mean values at different angles. The angle corresponding to the randomly oriented configuration ($\sim$$50~\mathrm{mrad}$) is not to scale. \textit{Right}: Mean \edep by photons as a function of their energy, at different values of  $\theta_{\text{mis}}$. Curves from the corresponding simulations (dashed lines) and extrapolations to $120~\mathrm{GeV}$ (squares) are also shown for the randomly oriented and axial cases. The extrapolations have been computed by fitting the experimental data to a logarithmic function.}
    \label{fig:photons}
\end{figure*}

Figure~\ref{fig:photons} (left) shows the distribution of the energy deposited in the crystal by $120~\mathrm{GeV}$ electrons for different values of $\theta_{\text{mis}}$. As expected, the deposited energy \edep decreases with increasing angle. Given the relatively small thickness of the crystal, the shower does not have sufficient time to fully develop, and the deposited energy is on the order of only tens to hundreds of MeV. As in the case of the thicker crystal, the deposited energy \edep reaches a maximum near $\theta_{\rm mis} = 0$ and decreases with increasing angle, with a measurable enhancement still observable beyond $0.5^\circ$.

Figure~\ref{fig:photons} (right) shows the average energy deposit inside the same crystal sample by the photon beam as a function of the inferred photon energy $E_\gamma^\mathrm{inf}$ (computed as explained in the \textit{Materials and methods} section) for different values of $\theta_{\text{mis}}$: on-axis, $0.9~\mathrm{mrad}$ ($1~\Theta_0$), $1.8~\mathrm{mrad}$ ($2~\Theta_0$) and random ($\sim 50~\mathrm{mrad}$). The energy deposited on axis grows significantly with $E_\gamma^\mathrm{inf}$ up to the maximum photon energy achievable with our setup, i.e., $100~\mathrm{GeV}$. There, the on-axis measurement shows a twofold enhancement in the deposited energy. At angles of $1~\Theta_0$ and $2~\Theta_0$, the enhancement in $E_{\text{dep}}$ is approximately $50\%$ and $25\%$, respectively. The extrapolated values at $120~\mathrm{GeV}$ are in good agreement with those obtained with $120~\mathrm{GeV}$ electrons, shown in figure~\ref{fig:photons} (left). It is also worth noting that a smaller axis-to-random enhancement was observed even at $E_\gamma^\mathrm{inf} \lesssim 25~\mathrm{GeV}$, i.e., below the SF threshold ($\chi < 1$). Overall, all the results are in excellent agreement with the predictions of the simulation model for both the on-axis and random cases, as shown by the dashed lines in figure~\ref{fig:photons} (right).

\begin{figure}[htbp!]
    \centering
    \includegraphics[width=\linewidth]{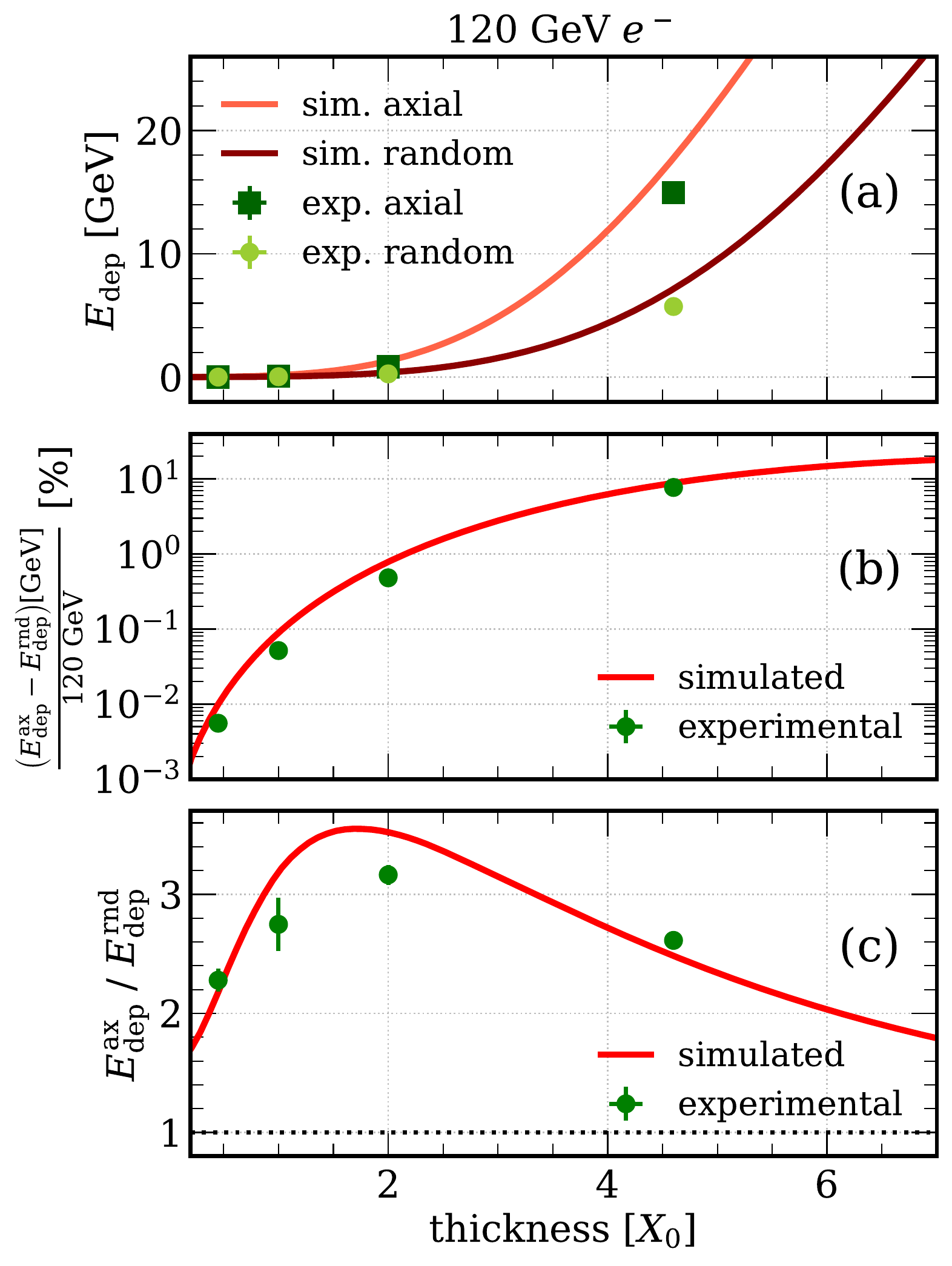}
    \caption{Measurements performed with the $120~\mathrm{GeV}$ electron beam as a function of the crystal thickness. \textit{(a)} Mean \edep in axial and random alignment. \textit{(b)} Difference between \edeprnd and \edepax normalized to the beam energy. \textit{(c)} Ratio between \edeprnd and \edepaxnospace. The solid curves were obtained with simulations.}
    \label{fig:showerdev}
\end{figure}

\subsection*{Electromagnetic shower development}
Finally, we exploited the rich set of experimental data collected with the $120~\mathrm{GeV}$ electron beam to investigate the features of shower development as a function of crystal thickness, ranging from $0.45~X_0$ to $4.6~X_0$. This allowed us to study the modification of electromagnetic shower evolution within the first few radiation lengths, where the SF effect is expected to be most pronounced. 

The mean energy deposit measured in the random (\edeprndnospace, squares) and axial (\edepaxnospace, circles) configurations as a function of the sample thickness is shown in figure \ref{fig:showerdev}a . Both \edeprnd and \edepax increase with crystal thickness, ranging from a few $\mathrm{MeV}$ for crystals less than $1~X_0$ thick, to the multi-$\mathrm{GeV}$ scale for the $4.6~X_0$ sample. \edepax increases more rapidly than \edeprnd, reaching higher values within the first few radiation lengths, providing clear evidence of the faster shower development. As demonstrated in the previous section, the MC simulations reproduce the experimental results well, allowing us to investigate how the deposited energy varies continuously with crystal depth, and to extrapolate it to depths larger than $4.6~X_0$.

This investigation becomes quite instructive if we focus on the difference between \edepax and \edeprndnospace, which strongly depends on the crystal thickness, and gives an indication on how a calorimetric measurement could be modified if the crystal is oriented. Figure~\ref{fig:showerdev}b displays the difference between \edeprnd and \edepax normalized to the beam energy vs. crystal depth. At a thickness of $1~X_0$, this difference accounts for a very small fraction of the beam energy (less than $0.05\%$). However, it increases significantly with thickness, reaching approximately $8\%$ at $4.6~X_0$.

Figure~\ref{fig:showerdev}c shows the ratio between $E_{\text{dep}}^{\text{axis}}$ and $E_{\text{dep}}^{\text{rnd}}$, providing an estimate of the enhancement in energy deposition observed on-axis compared to in random orientation. The maximum enhancement was observed at a depth of $2~X_0$, and it corresponds to a factor of approximately $3.2$. The experimental data are compared with simulated values of the fraction of primary energy deposited per radiation length as a function of the penetration depth in an ideal, very thick crystal ($> 20~X_0$). These simulations indicate an enhancement peak of approximately $3.6$ reached at a depth of about $1.7~X_0$, beyond which the curve decreases monotonically. This reduction in the enhancement, which still exceeds a factor of $2$ at $6~X_0$, can be attributed to the decreasing intensity of SF effects on secondary particles as their average energy diminishes. Additionally, their angular spread relative to the crystal axis may exceed $\Theta_0$. As a result, at more advanced stages of the shower development, the full SF regime is maintained only for a subset of particles, while the majority of the cascade evolves similarly to that in a randomly oriented crystal. This effect is clearly seen also in figure~\ref{fig:showermesh}, which illustrates the simulated energy deposition by $120$~$\mathrm{GeV}/c$ electrons in randomly (top) and axially (bottom) oriented PbWO$_4$ as a function of depth. 


\begin{figure*}[htbp!]
    \centering
    \includegraphics[width=0.9\linewidth]{mesh_random_fin.jpg}
    \includegraphics[width=0.9\linewidth]{mesh_axial_fin.jpg}
    \caption{Simulated energy density deposited in an electromagnetic shower initiated by $120$-$\mathrm{GeV}$ electrons as a function of position inside a randomly \textit{(top)} and axially \textit{(bottom)} oriented PbWO$_4$ crystal.}
    \label{fig:showermesh}
\end{figure*}

\section*{Discussion}

The observed enhancement in electromagnetic shower development, along with the associated increase in scintillation light yield during the early shower stages, may influence the interpretation of data from homogeneous crystal calorimeters, as well as their design, particularly in scenarios where incident particles are inadvertently aligned with the lattice axes. This phenomenon is relevant to existing detectors, as commonly used scintillating crystals are typically grown at a small angle relative to a principal axis, as is the case for the CMS ECAL~\cite{17_lecoq}. Such alignment effects may be especially significant in calorimeters with longitudinal segmentation, including space-based $\gamma$- and cosmic-ray detectors like the Fermi LAT~\cite{Atwood09} and the CALET calorimeter~\cite{calet}, which feature segmentation pitches of $1$–$2~X_0$. Lattice orientation will likewise be a critical factor in future high-energy experiments such as FCC-$ee$~\cite{IDEA2020} and the Muon Collider~\cite{crilin}, where crystal-based calorimeters with longitudinal segments of 4--5~$X_0$ are being considered.

Our findings could play a significant role in shaping the development of next-generation electromagnetic calorimeters, particularly those that may be designed to leverage the SF effects at high energies. In forward-geometry configurations, the enhancement in the containment of showers within the active volume would lead to an improved energy resolution. This is especially important in the first layer of a longitudinally segmented calorimeter with a small pitch (few-$X_0$), as shown in Figure~\ref{fig:showerdev}c. At higher energies, where the SF effects become more pronounced, this improvement is even more substantial.

These insights would also make it possible to design thinner calorimeters without sacrificing performance, a highly desirable feature for satellite-borne $\gamma$-ray telescopes, where constraints on the detector mass and volume are critical. A thinner design would allow for larger transverse dimensions, thus increasing the geometric acceptance---an important advantage for detecting rare, high-energy $\gamma$-rays.

Moreover, these features could also be beneficial to fixed-target experiments and forward calorimeters in collider experiments. Clearly, the enhanced shower containment would boost the detection efficiency of high-energy $\gamma$-rays, or guarantee the same efficiency of the current state of the art with a thinner design. At the same time, owing to the fact that hadronic interactions are largely unaffected by the lattice orientation, a compact, oriented crystal calorimeter would be more transparent to the passage of hadrons, which would meet the need of a highly efficient photon detector operating in a environment characterized by large neutral-hadron background \cite{hike_loi, hike_proposal_1_2}. Furthermore, if segmented, this design would feature improved $\gamma$/hadron discrimination, a critical feature for several forward physics and space-based experiments.

Finally, SF effects can be exploited in the search for feebly-interacting particles (FIPs), including potential light dark matter candidates, in beam dump/missing energy experiments such as NA64 at the CERN SPS~\cite{NA64}. In NA64 an electron beam of energy $100$--$150~\mathrm{GeV}$ is directed onto an electromagnetic calorimeter. Long-lived FIPs may be produced when the primary beam interacts in this detector. These FIPs do not interact with the calorimeter but can be detected indirectly by the missing-energy signature corresponding to the energy they carry away. The sensitivity of this approach is directly limited by the total length of the detector, since a FIP produced in its volume can only carry away all of its energy if it survives long enough to exit the calorimeter before decaying. A shorter calorimeter would therefore provide better sensitivity for FIPs with shorter lifetimes.

\section*{Conclusions}
In summary, we have reported detailed measurements of electromagnetic shower development and scintillation yield in axially oriented PbWO$_4$ crystals with thicknesses ranging from $0.45$ to $4.6~X_0$. How the development of showers is enhanced under axial alignment, relative to random orientation, was fully investigated and critically compared with MC simulations. Notably, the enhancement persists over a broad angular range, remaining measurable up to $1^\circ$. As the test samples closely resemble crystals used in existing and planned high-energy and $\gamma$-ray detectors, these results underscore the importance of accounting for lattice structure in calorimeter design---an aspect completely overlooked by the HEP community. Homogeneous crystal calorimeters provide the best energy resolution at the energy scales for HEP experiments, and are therefore essential elements of many existing and planned detectors. By incorporating the insights into the importance of scintillator crystal structure and orientation discussed here, the performance of future detectors can be significantly improved across various experimental domains.

\appendix
\section*{Acknowledgements}
This work was primarily funded by INFN CSN5 through the STORM project. We also acknowledge partial support of INFN CSN5 (OREO and Geant4-INFN projects) and CSN1 (NA62 experiment; RD-FLAVOUR project), of the Italian Ministry of University and Research (PRIN 2022Y87K7X) and of the European Commission (Horizon 2020 AIDAinnova, GA 101004761; Horizon 2020 MSCA IF Global TRILLION, GA 101032975).

We thank CRYTUR, spol. s.r.o. (Turnov, Czech Republic) and Molecular Technology (MolTech) GmbH (Berlin, Germany) for providing the crystals. Moreover, we thank the CERN PS/SPS coordinator and the SPS North Area staff for their support in the setup preparation: in particular, we are indebited to P. Boisseau-Bourgeois, S. Girod, M. Lazzaroni and B. Rae. Lastly, we warmly thank A. Zambonini for providing his proficient assistance in enhancing the visual elements of the manuscript.

\bibliographystyle{apsrev4-1}
\bibliography{bibliography}

\begin{thebibliography}{51}%
\makeatletter
\providecommand \@ifxundefined [1]{%
 \@ifx{#1\undefined}
}%
\providecommand \@ifnum [1]{%
 \ifnum #1\expandafter \@firstoftwo
 \else \expandafter \@secondoftwo
 \fi
}%
\providecommand \@ifx [1]{%
 \ifx #1\expandafter \@firstoftwo
 \else \expandafter \@secondoftwo
 \fi
}%
\providecommand \natexlab [1]{#1}%
\providecommand \enquote  [1]{``#1''}%
\providecommand \bibnamefont  [1]{#1}%
\providecommand \bibfnamefont [1]{#1}%
\providecommand \citenamefont [1]{#1}%
\providecommand \href@noop [0]{\@secondoftwo}%
\providecommand \href [0]{\begingroup \@sanitize@url \@href}%
\providecommand \@href[1]{\@@startlink{#1}\@@href}%
\providecommand \@@href[1]{\endgroup#1\@@endlink}%
\providecommand \@sanitize@url [0]{\catcode `\\12\catcode `\$12\catcode `\&12\catcode `\#12\catcode `\^12\catcode `\_12\catcode `\%12\relax}%
\providecommand \@@startlink[1]{}%
\providecommand \@@endlink[0]{}%
\providecommand \url  [0]{\begingroup\@sanitize@url \@url }%
\providecommand \@url [1]{\endgroup\@href {#1}{\urlprefix }}%
\providecommand \urlprefix  [0]{URL }%
\providecommand \Eprint [0]{\href }%
\providecommand \doibase [0]{http://dx.doi.org/}%
\providecommand \selectlanguage [0]{\@gobble}%
\providecommand \bibinfo  [0]{\@secondoftwo}%
\providecommand \bibfield  [0]{\@secondoftwo}%
\providecommand \translation [1]{[#1]}%
\providecommand \BibitemOpen [0]{}%
\providecommand \bibitemStop [0]{}%
\providecommand \bibitemNoStop [0]{.\EOS\space}%
\providecommand \EOS [0]{\spacefactor3000\relax}%
\providecommand \BibitemShut  [1]{\csname bibitem#1\endcsname}%
\let\auto@bib@innerbib\@empty
\bibitem [{\citenamefont {Fabjan}\ and\ \citenamefont {Gianotti}(2003)}]{gianotti2003}%
  \BibitemOpen
  \bibfield  {author} {\bibinfo {author} {\bibfnamefont {C.~W.}\ \bibnamefont {Fabjan}}\ and\ \bibinfo {author} {\bibfnamefont {F.}~\bibnamefont {Gianotti}},\ }\href {\doibase 10.1103/RevModPhys.75.1243} {\bibfield  {journal} {\bibinfo  {journal} {Rev. Mod. Phys.}\ }\textbf {\bibinfo {volume} {75}},\ \bibinfo {pages} {1243} (\bibinfo {year} {2003})}\BibitemShut {NoStop}%
\bibitem [{\citenamefont {Baier}\ \emph {et~al.}(1998)\citenamefont {Baier}, \citenamefont {Katkov},\ and\ \citenamefont {Strakhovenko}}]{baier1998electromagnetic}%
  \BibitemOpen
  \bibfield  {author} {\bibinfo {author} {\bibfnamefont {V.~N.}\ \bibnamefont {Baier}}, \bibinfo {author} {\bibfnamefont {V.~M.}\ \bibnamefont {Katkov}}, \ and\ \bibinfo {author} {\bibfnamefont {V.~V.}\ \bibnamefont {Strakhovenko}},\ }\href {\doibase 10.1142/2216} {\emph {\bibinfo {title} {Electromagnetic processes at high energies in oriented single crystals}}}\ (\bibinfo  {publisher} {World Scientific},\ \bibinfo {year} {1998})\BibitemShut {NoStop}%
\bibitem [{\citenamefont {Uggerh\o{}j}(2005)}]{uggerhoj2005}%
  \BibitemOpen
  \bibfield  {author} {\bibinfo {author} {\bibfnamefont {U.~I.}\ \bibnamefont {Uggerh\o{}j}},\ }\href {\doibase 10.1103/RevModPhys.77.1131} {\bibfield  {journal} {\bibinfo  {journal} {Rev. Mod. Phys.}\ }\textbf {\bibinfo {volume} {77}},\ \bibinfo {pages} {1131} (\bibinfo {year} {2005})}\BibitemShut {NoStop}%
\bibitem [{\citenamefont {Baryshevskii}\ and\ \citenamefont {Tikhomirov}(1989)}]{Baryshevskii89}%
  \BibitemOpen
  \bibfield  {author} {\bibinfo {author} {\bibfnamefont {V.~G.}\ \bibnamefont {Baryshevskii}}\ and\ \bibinfo {author} {\bibfnamefont {V.~V.}\ \bibnamefont {Tikhomirov}},\ }\href {\doibase 10.1070/PU1989v032n11ABEH002778} {\bibfield  {journal} {\bibinfo  {journal} {Sov. Phys. Usp.}\ }\textbf {\bibinfo {volume} {32}},\ \bibinfo {pages} {1013} (\bibinfo {year} {1989})}\BibitemShut {NoStop}%
\bibitem [{\citenamefont {S\o{}rensen}(1996)}]{sorensen1996_notes}%
  \BibitemOpen
  \bibfield  {author} {\bibinfo {author} {\bibfnamefont {A.~H.}\ \bibnamefont {S\o{}rensen}},\ }\href {\doibase 10.1016/0168-583X(96)00349-7} {\bibfield  {journal} {\bibinfo  {journal} {Nucl. Instrum. Methods Phys. Res. B}\ }\textbf {\bibinfo {volume} {119}},\ \bibinfo {pages} {2} (\bibinfo {year} {1996})}\BibitemShut {NoStop}%
\bibitem [{\citenamefont {Schwinger}(1951)}]{schwinger51}%
  \BibitemOpen
  \bibfield  {author} {\bibinfo {author} {\bibfnamefont {J.}~\bibnamefont {Schwinger}},\ }\href {\doibase 10.1103/PhysRev.82.664} {\bibfield  {journal} {\bibinfo  {journal} {Phys. Rev.}\ }\textbf {\bibinfo {volume} {82}},\ \bibinfo {pages} {664} (\bibinfo {year} {1951})}\BibitemShut {NoStop}%
\bibitem [{\citenamefont {Kimball}\ and\ \citenamefont {Cue}(1985)}]{kimball1985}%
  \BibitemOpen
  \bibfield  {author} {\bibinfo {author} {\bibfnamefont {J.~C.}\ \bibnamefont {Kimball}}\ and\ \bibinfo {author} {\bibfnamefont {N.}~\bibnamefont {Cue}},\ }\href {\doibase 10.1016/0370-1573(85)90021-3} {\bibfield  {journal} {\bibinfo  {journal} {Phys. Rep.}\ }\textbf {\bibinfo {volume} {125}},\ \bibinfo {pages} {69} (\bibinfo {year} {1985})}\BibitemShut {NoStop}%
\bibitem [{\citenamefont {Baryshevskii}\ and\ \citenamefont {Tikhomirov}(1983)}]{Baryshevskii:1983JETP}%
  \BibitemOpen
  \bibfield  {author} {\bibinfo {author} {\bibfnamefont {V.~G.}\ \bibnamefont {Baryshevskii}}\ and\ \bibinfo {author} {\bibfnamefont {V.~V.}\ \bibnamefont {Tikhomirov}},\ }\href@noop {} {\bibfield  {journal} {\bibinfo  {journal} {Sov. Phys. JETP}\ }\textbf {\bibinfo {volume} {58}},\ \bibinfo {pages} {135} (\bibinfo {year} {1983})}\BibitemShut {NoStop}%
\bibitem [{\citenamefont {Baryshevskii}\ and\ \citenamefont {Tikhomirov}(1985)}]{BARYSHEVSKII1985335}%
  \BibitemOpen
  \bibfield  {author} {\bibinfo {author} {\bibfnamefont {V.~G.}\ \bibnamefont {Baryshevskii}}\ and\ \bibinfo {author} {\bibfnamefont {V.~V.}\ \bibnamefont {Tikhomirov}},\ }\href {\doibase https://doi.org/10.1016/0375-9601(85)90178-1} {\bibfield  {journal} {\bibinfo  {journal} {Phys. Lett. A}\ }\textbf {\bibinfo {volume} {113}},\ \bibinfo {pages} {335} (\bibinfo {year} {1985})}\BibitemShut {NoStop}%
\bibitem [{\citenamefont {Baryshevsky}\ \emph {et~al.}(2017)\citenamefont {Baryshevsky} \emph {et~al.}}]{baryshevsky2017}%
  \BibitemOpen
  \bibfield  {author} {\bibinfo {author} {\bibfnamefont {V.}~\bibnamefont {Baryshevsky}} \emph {et~al.},\ }\href {\doibase 10.1016/j.nimb.2017.02.066} {\bibfield  {journal} {\bibinfo  {journal} {Nucl. Instrum. Methods Phys. Res. B}\ }\textbf {\bibinfo {volume} {402}},\ \bibinfo {pages} {35} (\bibinfo {year} {2017})}\BibitemShut {NoStop}%
\bibitem [{\citenamefont {{Particle Data Group}}(2022)}]{pdg}%
  \BibitemOpen
  \bibfield  {author} {\bibinfo {author} {\bibnamefont {{Particle Data Group}}},\ }\href {\doibase 10.1093/ptep/ptac097} {\bibfield  {journal} {\bibinfo  {journal} {Prog. Theor. Exp. Phys.}\ }\textbf {\bibinfo {volume} {2022}},\ \bibinfo {pages} {083C01} (\bibinfo {year} {2022})}\BibitemShut {NoStop}%
\bibitem [{\citenamefont {{CMS Collaboration}}(1997)}]{CMS_ECAL}%
  \BibitemOpen
  \bibfield  {author} {\bibinfo {author} {\bibnamefont {{CMS Collaboration}}},\ }\href@noop {} {\emph {\bibinfo {title} {{The CMS electromagnetic calorimeter project}: {Technical Design Report}}}},\ \bibinfo {type} {Tech. Rep.}\ (\bibinfo {year} {1997})\ \bibinfo {note} {{CERN-LHCC-97-033, CMS-TDR-4, CERN-LHCC-97-033, CMS-TDR-4}}\BibitemShut {NoStop}%
\bibitem [{\citenamefont {Erni}\ \emph {et~al.}(2008)\citenamefont {Erni} \emph {et~al.}}]{PANDA_ECAL}%
  \BibitemOpen
  \bibfield  {author} {\bibinfo {author} {\bibfnamefont {W.}~\bibnamefont {Erni}} \emph {et~al.} (\bibinfo {collaboration} {PANDA Collaboration}),\ }\href@noop {} {\emph {\bibinfo {title} {{Technical Design Report for PANDA Electromagnetic Calorimeter (EMC)}}}},\ \bibinfo {type} {Tech. Rep.}\ (\bibinfo {year} {2008})\ \Eprint {http://arxiv.org/abs/0810.1216} {arXiv:0810.1216 [physics.ins-det]} \BibitemShut {NoStop}%
\bibitem [{\citenamefont {Abbrescia}\ \emph {et~al.}(2025)\citenamefont {Abbrescia} \emph {et~al.}}]{FCC_IDEA}%
  \BibitemOpen
  \bibfield  {author} {\bibinfo {author} {\bibfnamefont {M.}~\bibnamefont {Abbrescia}} \emph {et~al.} (\bibinfo {collaboration} {IDEA Study Group}),\ }\href {https://cds.cern.ch/record/2926782} {\emph {\bibinfo {title} {{The IDEA detector concept for FCC-ee}}}},\ \bibinfo {type} {Tech. Rep.}\ (\bibinfo {year} {2025})\ \bibinfo {note} {{FERMILAB-PUB-25-0189-PPD}},\ \Eprint {http://arxiv.org/abs/2502.21223} {arXiv:2502.21223} \BibitemShut {NoStop}%
\bibitem [{\citenamefont {Soldani}(2023)}]{soldanithesis}%
  \BibitemOpen
  \bibfield  {author} {\bibinfo {author} {\bibfnamefont {M.}~\bibnamefont {Soldani}},\ }\emph {\bibinfo {title} {{Innovative applications of strong crystalline field effects to particle accelerators and detectors}}},\ \href {https://cds.cern.ch/record/2864634} {Ph.D. thesis},\ \bibinfo  {school} {Università degli Studi di Ferrara} (\bibinfo {year} {2023})\BibitemShut {NoStop}%
\bibitem [{\citenamefont {Dyson}\ and\ \citenamefont {\"Uberall}(1955)}]{dyson_anisotropy_1955}%
  \BibitemOpen
  \bibfield  {author} {\bibinfo {author} {\bibfnamefont {F.~J.}\ \bibnamefont {Dyson}}\ and\ \bibinfo {author} {\bibfnamefont {H.}~\bibnamefont {\"Uberall}},\ }\href {\doibase 10.1103/PhysRev.99.604} {\bibfield  {journal} {\bibinfo  {journal} {Phys. Rev.}\ }\textbf {\bibinfo {volume} {99}},\ \bibinfo {pages} {604} (\bibinfo {year} {1955})}\BibitemShut {NoStop}%
\bibitem [{\citenamefont {Diambrini-Palazzi}(1962)}]{diambrini-palazzi_interazioni_1962}%
  \BibitemOpen
  \bibfield  {author} {\bibinfo {author} {\bibfnamefont {G.}~\bibnamefont {Diambrini-Palazzi}},\ }\href {\doibase 10.1007/BF02860173} {\bibfield  {journal} {\bibinfo  {journal} {Nuovo Cim.}\ }\textbf {\bibinfo {volume} {25}},\ \bibinfo {pages} {88} (\bibinfo {year} {1962})}\BibitemShut {NoStop}%
\bibitem [{\citenamefont {Andersen}\ \emph {et~al.}(1983)\citenamefont {Andersen} \emph {et~al.}}]{andersen1983_notes}%
  \BibitemOpen
  \bibfield  {author} {\bibinfo {author} {\bibfnamefont {J.~U.}\ \bibnamefont {Andersen}} \emph {et~al.},\ }\href {\doibase 10.1146/annurev.ns.33.120183.002321} {\bibfield  {journal} {\bibinfo  {journal} {Annu. Rev. Nucl. Part.}\ }\textbf {\bibinfo {volume} {33}},\ \bibinfo {pages} {453} (\bibinfo {year} {1983})}\BibitemShut {NoStop}%
\bibitem [{\citenamefont {Cue}\ \emph {et~al.}(1984)\citenamefont {Cue} \emph {et~al.}}]{cue1984}%
  \BibitemOpen
  \bibfield  {author} {\bibinfo {author} {\bibfnamefont {N.}~\bibnamefont {Cue}} \emph {et~al.},\ }\href {\doibase 10.1103/PhysRevLett.53.972} {\bibfield  {journal} {\bibinfo  {journal} {Phys. Rev. Lett.}\ }\textbf {\bibinfo {volume} {53}},\ \bibinfo {pages} {972} (\bibinfo {year} {1984})}\BibitemShut {NoStop}%
\bibitem [{\citenamefont {Belkacem}\ \emph {et~al.}(1984)\citenamefont {Belkacem} \emph {et~al.}}]{belkacem1984}%
  \BibitemOpen
  \bibfield  {author} {\bibinfo {author} {\bibfnamefont {A.}~\bibnamefont {Belkacem}} \emph {et~al.},\ }\href {\doibase 10.1103/PhysRevLett.53.2371} {\bibfield  {journal} {\bibinfo  {journal} {Phys. Rev. Lett.}\ }\textbf {\bibinfo {volume} {53}},\ \bibinfo {pages} {2371} (\bibinfo {year} {1984})}\BibitemShut {NoStop}%
\bibitem [{\citenamefont {Moore}\ \emph {et~al.}(1996)\citenamefont {Moore} \emph {et~al.}}]{moore1996}%
  \BibitemOpen
  \bibfield  {author} {\bibinfo {author} {\bibfnamefont {R.}~\bibnamefont {Moore}} \emph {et~al.},\ }\href {\doibase 10.1016/0168-583X(96)00347-3} {\bibfield  {journal} {\bibinfo  {journal} {Nucl. Instrum. Methods Phys. Res. B}\ }\textbf {\bibinfo {volume} {119}},\ \bibinfo {pages} {149} (\bibinfo {year} {1996})}\BibitemShut {NoStop}%
\bibitem [{\citenamefont {Kirsebom}\ \emph {et~al.}(1998)\citenamefont {Kirsebom} \emph {et~al.}}]{kirsebom1998}%
  \BibitemOpen
  \bibfield  {author} {\bibinfo {author} {\bibfnamefont {K.}~\bibnamefont {Kirsebom}} \emph {et~al.},\ }\href {\doibase https://doi.org/10.1016/S0168-583X(97)00589-2} {\bibfield  {journal} {\bibinfo  {journal} {Nucl. Instrum. Methods Phys. Res. B}\ }\textbf {\bibinfo {volume} {135}},\ \bibinfo {pages} {143} (\bibinfo {year} {1998})}\BibitemShut {NoStop}%
\bibitem [{\citenamefont {Artru}\ \emph {et~al.}(2005)\citenamefont {Artru} \emph {et~al.}}]{artru2005}%
  \BibitemOpen
  \bibfield  {author} {\bibinfo {author} {\bibfnamefont {X.}~\bibnamefont {Artru}} \emph {et~al.},\ }\href {\doibase 10.1016/j.nimb.2005.04.134} {\bibfield  {journal} {\bibinfo  {journal} {Nucl. Instrum. Methods Phys. Res. B}\ }\textbf {\bibinfo {volume} {240}},\ \bibinfo {pages} {762} (\bibinfo {year} {2005})}\BibitemShut {NoStop}%
\bibitem [{\citenamefont {Bandiera}\ \emph {et~al.}(2022)\citenamefont {Bandiera} \emph {et~al.}}]{bandiera2022}%
  \BibitemOpen
  \bibfield  {author} {\bibinfo {author} {\bibfnamefont {L.}~\bibnamefont {Bandiera}} \emph {et~al.},\ }\href {\doibase 10.1140/epjc/s10052-022-10666-6} {\bibfield  {journal} {\bibinfo  {journal} {Eur. Phys. J. C}\ }\textbf {\bibinfo {volume} {82}},\ \bibinfo {pages} {699} (\bibinfo {year} {2022})}\BibitemShut {NoStop}%
\bibitem [{\citenamefont {Chehab}\ \emph {et~al.}(2002)\citenamefont {Chehab} \emph {et~al.}}]{chehab2002}%
  \BibitemOpen
  \bibfield  {author} {\bibinfo {author} {\bibfnamefont {R.}~\bibnamefont {Chehab}} \emph {et~al.},\ }\href {\doibase https://doi.org/10.1016/S0370-2693(01)01395-8} {\bibfield  {journal} {\bibinfo  {journal} {Phys. Lett. B}\ }\textbf {\bibinfo {volume} {525}},\ \bibinfo {pages} {41} (\bibinfo {year} {2002})}\BibitemShut {NoStop}%
\bibitem [{\citenamefont {Soldani}\ \emph {et~al.}(2023)\citenamefont {Soldani} \emph {et~al.}}]{soldani2023}%
  \BibitemOpen
  \bibfield  {author} {\bibinfo {author} {\bibfnamefont {M.}~\bibnamefont {Soldani}} \emph {et~al.},\ }\href {\doibase 10.1140/epjc/s10052-023-11247-x} {\bibfield  {journal} {\bibinfo  {journal} {Eur. Phys. J. C}\ }\textbf {\bibinfo {volume} {83}},\ \bibinfo {pages} {101} (\bibinfo {year} {2023})}\BibitemShut {NoStop}%
\bibitem [{\citenamefont {Chaikovska}\ \emph {et~al.}(2022)\citenamefont {Chaikovska} \emph {et~al.}}]{Chaikovska_2022}%
  \BibitemOpen
  \bibfield  {author} {\bibinfo {author} {\bibfnamefont {I.}~\bibnamefont {Chaikovska}} \emph {et~al.},\ }\href {\doibase 10.1088/1748-0221/17/05/P05015} {\bibfield  {journal} {\bibinfo  {journal} {J. Instrum.}\ }\textbf {\bibinfo {volume} {17}},\ \bibinfo {pages} {P05015} (\bibinfo {year} {2022})}\BibitemShut {NoStop}%
\bibitem [{\citenamefont {Bandiera}\ \emph {et~al.}(2018)\citenamefont {Bandiera} \emph {et~al.}}]{bandiera2018}%
  \BibitemOpen
  \bibfield  {author} {\bibinfo {author} {\bibfnamefont {L.}~\bibnamefont {Bandiera}} \emph {et~al.},\ }\href {\doibase 10.1103/PhysRevLett.121.021603} {\bibfield  {journal} {\bibinfo  {journal} {Phys. Rev. Lett.}\ }\textbf {\bibinfo {volume} {121}},\ \bibinfo {pages} {021603} (\bibinfo {year} {2018})}\BibitemShut {NoStop}%
\bibitem [{\citenamefont {Baskov}\ \emph {et~al.}(1999)\citenamefont {Baskov} \emph {et~al.}}]{baskov1999}%
  \BibitemOpen
  \bibfield  {author} {\bibinfo {author} {\bibfnamefont {V.}~\bibnamefont {Baskov}} \emph {et~al.},\ }\href {\doibase 10.1016/S0370-2693(99)00444-X} {\bibfield  {journal} {\bibinfo  {journal} {Phys. Lett. B}\ }\textbf {\bibinfo {volume} {456}},\ \bibinfo {pages} {86} (\bibinfo {year} {1999})}\BibitemShut {NoStop}%
\bibitem [{\citenamefont {Banerjee}\ \emph {et~al.}(2021)\citenamefont {Banerjee} \emph {et~al.}}]{banerjee_north_2021}%
  \BibitemOpen
  \bibfield  {author} {\bibinfo {author} {\bibfnamefont {D.}~\bibnamefont {Banerjee}} \emph {et~al.},\ }\href {https://cds.cern.ch/record/2774716} {\emph {\bibinfo {title} {The {North} {Experimental} {Area} at the {CERN} {Super} {Proton} {Synchrotron}}}},\ \bibinfo {type} {Tech. Rep.}\ (\bibinfo {year} {2021})\ \bibinfo {note} {{CERN}-ACC-NOTE-2021-0015}\BibitemShut {NoStop}%
\bibitem [{\citenamefont {Soldani}\ \emph {et~al.}(2022)\citenamefont {Soldani} \emph {et~al.}}]{soldani2022_pwo}%
  \BibitemOpen
  \bibfield  {author} {\bibinfo {author} {\bibfnamefont {M.}~\bibnamefont {Soldani}} \emph {et~al.},\ }\href {\doibase 10.1088/1742-6596/2374/1/012112} {\bibfield  {journal} {\bibinfo  {journal} {J. Phys. Conf. Ser.}\ }\textbf {\bibinfo {volume} {2374}},\ \bibinfo {pages} {012112} (\bibinfo {year} {2022})}\BibitemShut {NoStop}%
\bibitem [{\citenamefont {Selmi}\ \emph {et~al.}(2023)\citenamefont {Selmi} \emph {et~al.}}]{selmi2023}%
  \BibitemOpen
  \bibfield  {author} {\bibinfo {author} {\bibfnamefont {A.}~\bibnamefont {Selmi}} \emph {et~al.},\ }\href {\doibase 10.1016/j.nima.2022.167948} {\bibfield  {journal} {\bibinfo  {journal} {Nucl. Instrum. Methods Phys. Res. A}\ }\textbf {\bibinfo {volume} {1048}},\ \bibinfo {pages} {167948} (\bibinfo {year} {2023})}\BibitemShut {NoStop}%
\bibitem [{\citenamefont {Atwood}\ \emph {et~al.}(2009)\citenamefont {Atwood} \emph {et~al.}}]{Atwood09}%
  \BibitemOpen
  \bibfield  {author} {\bibinfo {author} {\bibfnamefont {W.~B.}\ \bibnamefont {Atwood}} \emph {et~al.},\ }\href {\doibase 10.1088/0004-637X/697/2/1071} {\bibfield  {journal} {\bibinfo  {journal} {Astrophys. J.}\ }\textbf {\bibinfo {volume} {697}},\ \bibinfo {pages} {1071} (\bibinfo {year} {2009})}\BibitemShut {NoStop}%
\bibitem [{\citenamefont {Lucchini}\ \emph {et~al.}(2020)\citenamefont {Lucchini} \emph {et~al.}}]{IDEA2020}%
  \BibitemOpen
  \bibfield  {author} {\bibinfo {author} {\bibfnamefont {M.}~\bibnamefont {Lucchini}} \emph {et~al.},\ }\href {\doibase 10.1088/1748-0221/15/11/P11005} {\bibfield  {journal} {\bibinfo  {journal} {J. Instrum.}\ }\textbf {\bibinfo {volume} {15}},\ \bibinfo {pages} {P11005} (\bibinfo {year} {2020})}\BibitemShut {NoStop}%
\bibitem [{\citenamefont {Ceravolo}\ \emph {et~al.}(2022)\citenamefont {Ceravolo} \emph {et~al.}}]{crilin}%
  \BibitemOpen
  \bibfield  {author} {\bibinfo {author} {\bibfnamefont {S.}~\bibnamefont {Ceravolo}} \emph {et~al.},\ }\href {\doibase 10.1088/1748-0221/17/09/P09033} {\bibfield  {journal} {\bibinfo  {journal} {J. Instrum.}\ }\textbf {\bibinfo {volume} {17}},\ \bibinfo {pages} {P09033} (\bibinfo {year} {2022})}\BibitemShut {NoStop}%
\bibitem [{\citenamefont {Monti-Guarnieri}\ \emph {et~al.}(2022)\citenamefont {Monti-Guarnieri} \emph {et~al.}}]{pmg2022}%
  \BibitemOpen
  \bibfield  {author} {\bibinfo {author} {\bibfnamefont {P.}~\bibnamefont {Monti-Guarnieri}} \emph {et~al.},\ }\href {\doibase 10.22323/1.414.0342} {\bibfield  {journal} {\bibinfo  {journal} {PoS}\ }\textbf {\bibinfo {volume} {ICHEP2022}},\ \bibinfo {pages} {342} (\bibinfo {year} {2022})}\BibitemShut {NoStop}%
\bibitem [{\citenamefont {Bandiera}\ \emph {et~al.}(2013)\citenamefont {Bandiera} \emph {et~al.}}]{BANDIERA2013135}%
  \BibitemOpen
  \bibfield  {author} {\bibinfo {author} {\bibfnamefont {L.}~\bibnamefont {Bandiera}} \emph {et~al.},\ }\href {\doibase https://doi.org/10.1016/j.nimb.2013.02.029} {\bibfield  {journal} {\bibinfo  {journal} {Nucl. Instrum. Methods Phys. Res. B}\ }\textbf {\bibinfo {volume} {309}},\ \bibinfo {pages} {135} (\bibinfo {year} {2013})}\BibitemShut {NoStop}%
\bibitem [{\citenamefont {Lietti}\ \emph {et~al.}(2013)\citenamefont {Lietti} \emph {et~al.}}]{lietti2013}%
  \BibitemOpen
  \bibfield  {author} {\bibinfo {author} {\bibfnamefont {D.}~\bibnamefont {Lietti}} \emph {et~al.},\ }\href {\doibase 10.1016/j.nima.2013.07.066} {\bibfield  {journal} {\bibinfo  {journal} {Nucl. Instrum. Methods Phys. Res. A}\ }\textbf {\bibinfo {volume} {729}},\ \bibinfo {pages} {527} (\bibinfo {year} {2013})}\BibitemShut {NoStop}%
\bibitem [{\citenamefont {{OPAL Collaboration}}(1991)}]{opal}%
  \BibitemOpen
  \bibfield  {author} {\bibinfo {author} {\bibnamefont {{OPAL Collaboration}}},\ }\href {\doibase https://doi.org/10.1016/0168-9002(91)90547-4} {\bibfield  {journal} {\bibinfo  {journal} {Nucl. Instrum. Methods Phys. Res. A}\ }\textbf {\bibinfo {volume} {305}},\ \bibinfo {pages} {275} (\bibinfo {year} {1991})}\BibitemShut {NoStop}%
\bibitem [{\citenamefont {Monti-Guarnieri}(2023)}]{pmg2023}%
  \BibitemOpen
  \bibfield  {author} {\bibinfo {author} {\bibfnamefont {P.}~\bibnamefont {Monti-Guarnieri}},\ }\href {\doibase 10.1393/ncc/i2023-23098-5} {\bibfield  {journal} {\bibinfo  {journal} {Nuovo Cim. C}\ }\textbf {\bibinfo {volume} {46}},\ \bibinfo {pages} {98} (\bibinfo {year} {2023})}\BibitemShut {NoStop}%
\bibitem [{\citenamefont {Agostinelli}\ \emph {et~al.}(2003)\citenamefont {Agostinelli} \emph {et~al.}}]{g4}%
  \BibitemOpen
  \bibfield  {author} {\bibinfo {author} {\bibfnamefont {S.}~\bibnamefont {Agostinelli}} \emph {et~al.},\ }\href {\doibase https://doi.org/10.1016/S0168-9002(03)01368-8} {\bibfield  {journal} {\bibinfo  {journal} {Nucl. Instrum. Methods Phys. Res. A}\ }\textbf {\bibinfo {volume} {506}},\ \bibinfo {pages} {250} (\bibinfo {year} {2003})}\BibitemShut {NoStop}%
\bibitem [{\citenamefont {Guidi}\ \emph {et~al.}(2012)\citenamefont {Guidi}, \citenamefont {Bandiera},\ and\ \citenamefont {Tikhomirov}}]{Guidi12}%
  \BibitemOpen
  \bibfield  {author} {\bibinfo {author} {\bibfnamefont {V.}~\bibnamefont {Guidi}}, \bibinfo {author} {\bibfnamefont {L.}~\bibnamefont {Bandiera}}, \ and\ \bibinfo {author} {\bibfnamefont {V.}~\bibnamefont {Tikhomirov}},\ }\href {\doibase 10.1103/PhysRevA.86.042903} {\bibfield  {journal} {\bibinfo  {journal} {Phys. Rev. A}\ }\textbf {\bibinfo {volume} {86}},\ \bibinfo {pages} {042903} (\bibinfo {year} {2012})}\BibitemShut {NoStop}%
\bibitem [{\citenamefont {Bandiera}\ \emph {et~al.}(2015)\citenamefont {Bandiera}, \citenamefont {Bagli}, \citenamefont {Guidi},\ and\ \citenamefont {Tikhomirov}}]{RADCHARM++}%
  \BibitemOpen
  \bibfield  {author} {\bibinfo {author} {\bibfnamefont {L.}~\bibnamefont {Bandiera}}, \bibinfo {author} {\bibfnamefont {E.}~\bibnamefont {Bagli}}, \bibinfo {author} {\bibfnamefont {V.}~\bibnamefont {Guidi}}, \ and\ \bibinfo {author} {\bibfnamefont {V.~V.}\ \bibnamefont {Tikhomirov}},\ }\href {\doibase https://doi.org/10.1016/j.nimb.2015.03.031} {\bibfield  {journal} {\bibinfo  {journal} {Nucl. Instrum. Methods Phys. Res. B}\ }\textbf {\bibinfo {volume} {355}},\ \bibinfo {pages} {44} (\bibinfo {year} {2015})}\BibitemShut {NoStop}%
\bibitem [{\citenamefont {Sytov}\ \emph {et~al.}(2019)\citenamefont {Sytov}, \citenamefont {Tikhomirov},\ and\ \citenamefont {Bandiera}}]{CRYSTALRAD}%
  \BibitemOpen
  \bibfield  {author} {\bibinfo {author} {\bibfnamefont {A.~I.}\ \bibnamefont {Sytov}}, \bibinfo {author} {\bibfnamefont {V.~V.}\ \bibnamefont {Tikhomirov}}, \ and\ \bibinfo {author} {\bibfnamefont {L.}~\bibnamefont {Bandiera}},\ }\href {\doibase 10.1103/PhysRevAccelBeams.22.064601} {\bibfield  {journal} {\bibinfo  {journal} {Phys. Rev. Accel. Beams}\ }\textbf {\bibinfo {volume} {22}},\ \bibinfo {pages} {064601} (\bibinfo {year} {2019})}\BibitemShut {NoStop}%
\bibitem [{\citenamefont {Sytov}\ \emph {et~al.}(2023)\citenamefont {Sytov} \emph {et~al.}}]{sytov2023}%
  \BibitemOpen
  \bibfield  {author} {\bibinfo {author} {\bibfnamefont {A.}~\bibnamefont {Sytov}} \emph {et~al.},\ }\href {\doibase 10.1007/s40042-023-00834-6} {\bibfield  {journal} {\bibinfo  {journal} {J. Korean Phys. Soc.}\ }\textbf {\bibinfo {volume} {83}},\ \bibinfo {pages} {132} (\bibinfo {year} {2023})}\BibitemShut {NoStop}%
\bibitem [{\citenamefont {Bandiera}\ \emph {et~al.}(2021)\citenamefont {Bandiera} \emph {et~al.}}]{Bandiera2021}%
  \BibitemOpen
  \bibfield  {author} {\bibinfo {author} {\bibfnamefont {L.}~\bibnamefont {Bandiera}} \emph {et~al.},\ }\href {\doibase 10.1140/epjc/s10052-021-09071-2} {\bibfield  {journal} {\bibinfo  {journal} {Eur. Phys. J. C}\ }\textbf {\bibinfo {volume} {81}},\ \bibinfo {pages} {284} (\bibinfo {year} {2021})}\BibitemShut {NoStop}%
\bibitem [{\citenamefont {Asaoka}\ \emph {et~al.}(2017)\citenamefont {Asaoka} \emph {et~al.}}]{calet}%
  \BibitemOpen
  \bibfield  {author} {\bibinfo {author} {\bibfnamefont {Y.}~\bibnamefont {Asaoka}} \emph {et~al.},\ }\href {\doibase https://doi.org/10.1016/j.astropartphys.2017.03.002} {\bibfield  {journal} {\bibinfo  {journal} {Astropart. Phys.}\ }\textbf {\bibinfo {volume} {91}},\ \bibinfo {pages} {1} (\bibinfo {year} {2017})}\BibitemShut {NoStop}%
\bibitem [{\citenamefont {Lecoq}\ \emph {et~al.}(2017)\citenamefont {Lecoq}, \citenamefont {Getkin},\ and\ \citenamefont {Korzhik}}]{17_lecoq}%
  \BibitemOpen
  \bibfield  {author} {\bibinfo {author} {\bibfnamefont {P.}~\bibnamefont {Lecoq}}, \bibinfo {author} {\bibfnamefont {A.}~\bibnamefont {Getkin}}, \ and\ \bibinfo {author} {\bibfnamefont {M.}~\bibnamefont {Korzhik}},\ }\href {\doibase 10.1007/978-3-319-45522-8} {\emph {\bibinfo {title} {{Inorganic Scintillators for Detector Systems}}}}\ (\bibinfo  {publisher} {Springer},\ \bibinfo {year} {2017})\ \bibinfo {note} {{Chapter 8 and references therein}}\BibitemShut {NoStop}%
\bibitem [{\citenamefont {Cortina~Gil}\ \emph {et~al.}(2022)\citenamefont {Cortina~Gil} \emph {et~al.}}]{hike_loi}%
  \BibitemOpen
  \bibfield  {author} {\bibinfo {author} {\bibfnamefont {E.}~\bibnamefont {Cortina~Gil}} \emph {et~al.},\ }\href {https://cds.cern.ch/record/2839661} {\emph {\bibinfo {title} {{HIKE, High Intensity Kaon Experiments at the CERN SPS: Letter of Intent}}}},\ \bibinfo {type} {Tech. Rep.}\ (\bibinfo {year} {2022})\ \bibinfo {note} {{CERN}-SPSC-2022-031, SPSC-I-257, SPSC-I-257}\BibitemShut {NoStop}%
\bibitem [{\citenamefont {{HIKE Collaboration}}(2023)}]{hike_proposal_1_2}%
  \BibitemOpen
  \bibfield  {author} {\bibinfo {author} {\bibnamefont {{HIKE Collaboration}}},\ }\href {https://cds.cern.ch/record/2878543} {\emph {\bibinfo {title} {{High Intensity Kaon Experiments (HIKE) at the CERN SPS: Proposal for Phases 1 and 2}}}},\ \bibinfo {type} {Tech. Rep.}\ \bibinfo {number} {CERN-SPSC-2023-031, SPSC-P-368}\ (\bibinfo {year} {2023})\BibitemShut {NoStop}%
\bibitem [{\citenamefont {Andreev}\ \emph {et~al.}(2023)\citenamefont {Andreev} \emph {et~al.}}]{NA64}%
  \BibitemOpen
  \bibfield  {author} {\bibinfo {author} {\bibfnamefont {Y.~M.}\ \bibnamefont {Andreev}} \emph {et~al.} (\bibinfo {collaboration} {NA64 Collaboration}),\ }\href {\doibase 10.1103/PhysRevLett.131.161801} {\bibfield  {journal} {\bibinfo  {journal} {Phys. Rev. Lett.}\ }\textbf {\bibinfo {volume} {131}},\ \bibinfo {pages} {161801} (\bibinfo {year} {2023})}\BibitemShut {NoStop}%
\end{thebibliography}%
\end{document}